\documentclass[doublecol]{epl2}

\usepackage{graphicx}
\usepackage{dcolumn}% Align table columns on decimal point
\usepackage{bm}% bold math
\usepackage{color}

\title{ Conductance through quantum wires with L\'evy-type disorder:
universal statistics in anomalous quantum transport}
\shorttitle{Conductance through quantum wires with L\'evy-type disorder}
\author{F. Falceto \and V. A. Gopar \inst{}}
\institute{Depto de F\'isica
Te\'orica, Facultad de Ciencias, and Instituto de Biocomputaci\'on y
F\'isica de Sistemas Complejos (BIFI), Universidad de Zaragoza, Pedro
Cerbuna 12, E-50009, Zaragoza, Spain, EU}

\abstract{ In this letter we study the conductance $G$ through one-dimensional
quantum wires with disorder configurations characterized by long-tailed
distributions (L\'evy-type disorder). We calculate analytically the
conductance distribution which reveals a universal
statistics: the distribution of conductances is fully determined by the
exponent $\alpha$ of the power-law decay of the disorder distribution
and the average $\langle \ln G \rangle$, i.e., all other details of the
disorder configurations are irrelevant. For $0< \alpha < 1$ we found
that the fluctuations of $\ln G$ are not self-averaging and $\langle \ln
G \rangle$ scales with the length of the system as $L^\alpha$, in
contrast to the predictions of the standard scaling-theory of
localization where $\ln G$ is a self-averaging quantity and $\langle \ln
G \rangle$ scales linearly with $L$.  Our theoretical results are
verified by comparing with numerical simulations of one-dimensional
disordered wires.}

\pacs{72.10.-d}{Theory of electronic transport; scattering mechanisms}
\pacs{72.15.Rn}{Localization effects (Anderson or weak localization) }
\pacs{73.21.Hb}{Quantum wires}

\begin{document}
\maketitle

Quantum coherent electronic transport through disordered conductors has
been widely studied from a fundamental and practical point of view. For
instance, the pioneering ideas by Anderson on the localization of
electron wave functions in disordered conductors and the one-parameter
scaling approach to this phenomenon \cite{anderson} have been of large
impact on condensed matter physics. The progress in the theoretical
description of the electronic transport has been accompanied and
stimulated by advances in the fabrication of small electronic devices,
where coherent electronic transport is possible.

On the other hand, random processes characterized by probability
densities (L\'evy-type distributions) $\rho(x)$ with long tails, i.e.,
for large $x$, $\rho(x) \sim 1/x^{1+\alpha}$ with $0 <\alpha <2$, have
been found in very different phenomena in nature and human activities.
For instance, it has been seen that the movement patterns of some
marine predators follow a L\'evy-type distribution \cite{fishes}. In
economy, fluctuations of the stock market indices can be described by
L\'evy models \cite{stock}. Mathematicians have been investigating the
properties of this class of probability densities since seminal works by
L\'evy \cite{levy-kolmorogov-calvo, uchaikin}. Among these
properties we remark the generalized central limit theorem (GCLT). We
recall that the central limit theorem states that the normalized sum of
independent variables with finite mean and variance is normally
distributed in the limit of a large number of variables; the GCLT gets
rid of the finite variance condition and the resulting limit probability
distribution is a L\'evy or $\alpha$-stable distribution. In physics,
L\'evy-type distributions have been applied to several areas such as
statistical mechanics, fluids, and dynamics of chaotic systems
\cite{uchaikin}. Therefore L\'evy processes have become of a broad
interdisciplinary interest.

L\'evy processes have been found in different phenomena, but an
experimental study of these processes in a controllable way has not been
possible until very recently.  New techniques in the fabrication of
materials have allowed the experimental realization and study of L\'evy
transport \cite{hideo, barthelemy}. For instance, in Ref.
\cite{hideo} SiC nanowires have been fabricated via
self-organized processes. It turns out that the diameter of the SiC
nanowires shows fluctuations whose distribution follows a power-law
decay. In Ref. \cite{barthelemy} Barthelemy and collaborators  were able to make a
disordered medium where light travels across glass microspheres whose
diameter follows a L\'evy-type distribution.  Motivated by these kind of
experiments, here we shall study the quantum coherent transport of
electrons across one-dimensional (1D) disordered wires with L\'evy-type
disorder. Previous theoretical works
\cite{lambert-boose-raffaella-beenakker-malyshev} have pointed out and studied
the effects of L\'evy processes in the electronic transport.

In this work we obtain the distribution of the conductance of electrons
traveling through a 1D disordered wire whose configuration of impurities
follows a density distribution with a power-law tail, fig. 1. We assume
that the electrons travel coherently across the sample. Thus
interference effects are reflected in  transport quantities,
such as the conductance, and due to the L\'evy-type distribution of the
scatterers novel effects are seen. For instance, the coexistence of
ballistic and insulating regimes. More interesting from a fundamental
point of view is that a new universality class emerges, as the
distribution of the conductance is completely determined by the
power-law tail exponent $\alpha$ of the disorder distribution and the
ensemble average $\langle \ln G \rangle$. No further information is
needed to determined the conductance statistics.
\begin{figure}
\begin{center}
 \includegraphics[width=0.9\columnwidth]{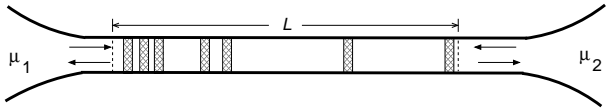}
\end{center}
 \caption{Schematic of a quantum wire with scatterers (shaded regions)
randomly placed accordingly to a L\'evy-type distribution. The wire is
fed with electrons by the reservoirs $\mu_{1, 2}$}
\end{figure}

To study the conductance we adopt the Landauer-B\"uttiker approach in
which the electronic transport is seen as a scattering problem. In this
approach the dimensionless conductance $G$ (in units of the conductance
quantum $2e^2/\hbar$) is given by the transmission coefficient.  With
respect to the statistical study of the conductance, we shall use some
results from the standard scaling theory developed in Refs. \cite{
melnikov-dorokhov, mello_groups}.  These works considered the weak
scattering limit ($\lambda_F \ll l$, being $\lambda_F$ and $l$ the Fermi
wavelength and the mean free-path, respectively) and found that the
evolution of the conductance distribution as a function of the length of
the 1D wire is governed by a Fokker-Planck equation. Thus the solution
of such a diffusion equation gives the distribution of conductances
$p_s(G)$ for a wire of length $L$, which is written as \cite{carlo-review}
\begin{equation}
\label{pofg}
p_s(G)=\frac{s^{-\frac{3}{2}}}{\sqrt{2\pi}} \frac{{\rm
e}^{-\frac{s}{4}}}{G^2}\int_{y_0}^{\infty}dy\frac{y{\rm
e}^{-\frac{y^2}{4s}}} {\sqrt{\cosh{y}+1-2/G}},
\end{equation}
where $y_0={\rm arcosh}{(2/G-1)}$ and $s=L/l$. We make a couple of
general comments on the scaling approach of Ref. \cite{mello_groups}. On
the one hand, in order to obtain the above-mentioned diffusion equation,
it is assumed that for a very small sample of length $\delta L$, the
resistance average is fixed and depends linearly on $\delta L$. As a
consequence the average $\langle - \ln G \rangle$ of the whole sample
scales linearly with $L$: $\langle -\ln G \rangle=L/l$. On the other
hand, one can see from eq. (\ref{pofg}) that the distribution of the
conductance is determined by the single parameter $s$. This is a
remarkable fact since it means that all the information we need to give
a statistical description of the conductance is the system length
measured in units of the mean-free path, all other microscopic details
are irrelevant. This point was referred as a central-limit theorem for
weakly disordered wires and eq. (\ref{pofg}) shows explicitly the
single parameter dependence of the statistics of the transport problem. In
relation to the previous comments, we shall see that for Levy-type
disorder $\langle \ln G \rangle $ does not scale linearly with $L$ but still
the conductance
distribution is determined by the average $\langle \ln G \rangle$ and
the exponent of the power-law decay of the disorder-configuration
distribution.

\begin{figure}
\begin{center}
 \includegraphics[width=0.9\columnwidth]{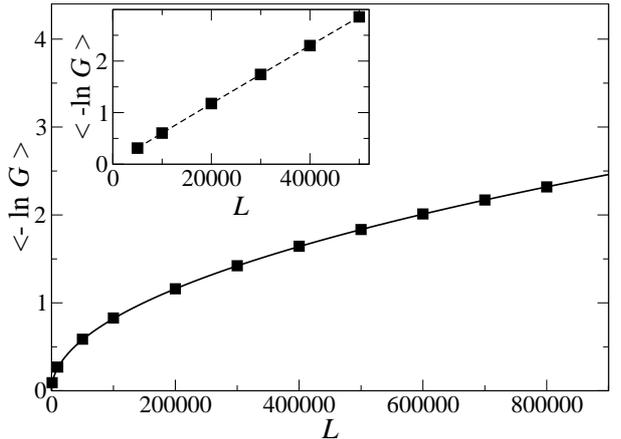}
\end{center}
\caption{The average $ \langle - \ln G \rangle$ as a function of $L$ for
a wire with potential barriers whose separations and widths are
generated accordingly to a Levy-type distribution with $\alpha=1/2$ (eq.
(\ref{asymptotics})\,). As we predict $\langle - \ln G \rangle \propto
L^{1/2}$. The solid line is obtained from eq. (\ref{lngofl_a}). A good
agreement is seen between numerics (squares) and theory. Inset: $
\langle - \ln G \rangle$ for a disorder distribution with $\alpha=3/2$.
In agreement with our model, $\langle -\ln G \rangle \propto L$, for
$\alpha \ge 1$, as in usual disorder systems.} \label{fig_2}
\end{figure}

As a further motivation, let us illustrate the fact that the ensemble
average $\langle \ln G \rangle$ does not scale linearly with $L$ in the
presence of long-range disorder for $\alpha <1$. In fig. \ref{fig_2} we
show the numerical results for $\langle - \ln G \rangle$ for two
different values of $\alpha$. For the numerical simulations, we consider
a succession of square-potentials barriers with fixed hight, but
separations and widths randomly generated from a Levy-type distribution
\cite{numericalgeneration}. The statistics is collected across an
ensemble of different disorder realizations where the length of the
system is fixed. For $\alpha=1/2$, we can see clearly from fig.
\ref{fig_2} (main frame) that $\langle - \ln G \rangle$ is not a linear
function of $L$, whereas for $\alpha =3/2$ (inset of fig. \ref{fig_2})
$\langle - \ln G \rangle \propto L$, as in the standard scaling
approach. The solid line in fig. \ref{fig_2} is obtained from our
theoretical model which we explain below.
 %%%%%%%%%
In order to describe the results shown in fig. \ref{fig_2} and calculate
the conductance distribution, it is essential to consider the strong
fluctuations in the number of scatterers in a wire of length $L$ due to
the long tails of the density distribution of the
disorder-configurations. For instance, if the tail of the density decays
slowly as
\begin{equation}
\label{asymptotics}
\rho(x)\sim \frac{c}{x^{1+\alpha}},
\end{equation}
for large $x$ and $0<\alpha< 1$, the first and second moment diverges.
Actually, we shall concentrate precisely in these values of $\alpha$ as
they exhibit the most interesting results for the conductance
statistics. For the case $1<\alpha <2$, where the first moment is
finite, the statistical properties of the conductance are similar to
those described by the standard scaling theory (e.g., see the inset of fig.
\ref{fig_2}). This situation can also be treated along similar lines of this work
\cite{elsewhere}.

Let us start by introducing the probability density $\Pi_{L}(\nu)$ for
the number of scattering-units $\nu$ in a wire of length $L$. If the
separations and widths of the scattering-units in the wire follow a law
like eq. (\ref{asymptotics}) one can show, using the properties of the
L\'evy distributions that
\footnote[1]{
Let us approximate the number of scatterers-units $n$ by a continuos variable $\nu$
such that the probability of having exactly $n$ scatterers in a system of length $L$ is
given by $\int^{n+1/2}_{n-1/2}\Pi_L(\nu){\rm d}\nu$. Therefore we have that
$$
\int_0^{n+1/2} \Pi_L(\nu) {\rm d}\nu=\int _L^\infty \rho_{2n+1}(x){\rm d}x ,
$$
where $\rho_m(x)$  is the mth auto-convolution of $\rho(x)$ and represents the PDF of the total
length after $m$ drawings.

Introducing the new variables $\overline \nu=L/(2\nu)^{1/\alpha}$, $\overline x=x/(2n+1)^{1/\alpha}$,  and $\overline\Pi_L(\overline\nu)=(\alpha/2)L^\alpha\overline\nu^{-\alpha-1}\Pi(L^\alpha\overline\nu^{-\alpha}/2)$. The
previous relation reads
$$
\int_A^\infty\overline{ \Pi}_L(\overline\nu) {\rm d}\overline\nu =
\int_A^\infty (2n+1)^{1/\alpha}\rho_{2n+1} ((2n+1)^{1/\alpha}\overline x) {\rm d}\overline x ,
$$
where $A=L(2n+1)^{-1/\alpha}$. If we take now the limit of large $L$ keeping $A$
constant and use the generalization of the central limit theorem which states that
$$\lim_{m\to\infty} m^{1/\alpha} \rho_m(m^{1/\alpha}\overline x)= q_{\alpha,c}(\overline x);$$
we obtain the relation
$$\lim_{L\to\infty}\int_A^\infty\overline\Pi_L(\overline\nu) {\rm d}\overline\nu
= \int_A^\infty  q_{\alpha,c}(\overline x) {\rm d}\overline{x},$$
from which our result,  Eq. (\ref{pi_alpha<1}),  follows.
}
\begin{equation}
\label{pi_alpha<1}
 \Pi_{L}(\nu)=\frac2\alpha\frac{L}{(2\nu)^{\frac{1+\alpha}{\alpha}}}
q_{\alpha,c}({L}/{(2\nu)^{1/\alpha}}) ,
\end{equation}
for $ 0 < \alpha <1 $, in the macroscopic limit $L\gg c^{1/\alpha}$.
Here $q_{\alpha,c}$ is the probability density function (PDF) of the
L\'evy distribution supported in the positive semiaxis that behaves like
$\rho(x)$ for large values of $x$ [see eq. (\ref{asymptotics})]. The
concrete form of $q_{\alpha,c}(x)$ is better expressed using the Fourier
transform
\begin{equation}\label{levy}
\widehat {q}_{\alpha,c}(k)= \exp \left(-|k|^\alpha\left( B\theta(k)+
B^*\theta(-k) \right)\right) ,
\end{equation}
where $\theta$ is the Heaviside step function and $B= -c
{\Gamma(-\alpha)} {\rm e}^{i\frac{\pi\alpha}2}$, $\Gamma$ being the
Gamma function. Note that the result for $\Pi_{L}(\nu)$, eq.
(\ref{pi_alpha<1}), depends only of the parameters $c$ and $\alpha$. All
other details of the PDF $\rho$ of the disorder configurations are
washed out. This fact follows from the generalization of the central
limit theorem to distributions with long tails (see footnote 1).

In order to continue we introduce some notation: let $\langle \ln
G\rangle_\nu$ and $\langle \ln G\rangle_L$ be the expectation values for
samples with a fixed number of scatterers $\nu$ and fixed length $L$,
respectively. From the standard scaling theory of localization $\langle
-\ln G\rangle_\nu$ is proportional to $\nu$: $\langle -\ln G\rangle_\nu
= a \nu$, $a$ being a constant. Thus we have that
\begin{eqnarray}
\label{lngofl}
\langle -\ln G\rangle_L&=&\int_0^\infty  \langle -\ln G\rangle_\nu  \Pi_{L}(\nu) {\rm d}\nu  \\
&=&\int_0^\infty a \nu \frac2\alpha\frac{L}{(2\nu)^{\frac{1+\alpha}{\alpha}}}
q_{\alpha,c}({L}/{(2\nu)^{1/\alpha}}) {\rm d}\nu ,
\end{eqnarray}
where we have used eq. (\ref{pi_alpha<1}). Using the scaling property of
the L\'evy distributions, namely
$c^{1/\alpha}q_{\alpha,c}(c^{1/\alpha}x)=q_{\alpha,1}(x)$ and making the
change of variable $ z=L/(2c\nu)^{1\over\alpha}, $ we finally obtain
\begin{equation} \label{lngofl_a} \langle -\ln G\rangle_L = a
\frac{L^\alpha}c \frac12\int_0^\infty z^{-\alpha}q_{\alpha,1}(z){\rm d}z
= a \frac{L^\alpha}{c} I_\alpha, \end{equation} where $I_\alpha =
(1/2)\int_0^\infty z^{-\alpha}q_{\alpha,1}(z){\rm d}z$.  The last
equality in (\ref{lngofl_a}) stands to emphasize the r\^ole played by
the different elements ($c, \alpha$) of the original distribution and
the physical characteristics ($a, L$) of the system. We have shown,
therefore, that $\langle -\ln G\rangle_L\propto L^\alpha$, in contrast
to the linear behavior with $L$ expected from the usual scaling theory.
This is true for any value of $0<\alpha<1$, while the standard linear
behavior is recovered for $\alpha\geq 1$. This was already illustrated
in the inset of fig. \ref{fig_2}. Moreover, $\ln G$ for $\alpha <1$ is
not self-averaging. That is, in the limit of large samples we found that
the ratio $R= \lim_{L \to \infty} \mathrm{var}(\ln G)/\langle \ln G
\rangle^2 $ is non zero for $\alpha \le 1$, whereas $R = 0$ for $\alpha >
1$, as in the conventional scaling theory.

We now calculate the conductance distribution. As we mentioned before,
in the standard scaling theory the complete conductance statistics is fully
determined  $\langle \ln G\rangle_L$. We shall
show that the same is true when the configuration of scattering-units
follows a PDF with long tails, with the only additional information of
the parameter $\alpha$ that characterizes the distribution tail.

Assume that we have an ensemble of disordered wires with $\langle -\ln G
\rangle_L=\xi$ whose scattering-units are distributed according to a
long-tailed probability, as described before. The conductance
distribution $P_\xi(G)$ is given by an integral that involves $p_s(G)$,
given in eq. ($\ref{pofg}$) with the parameter $s$ related to the number
of scatterers, $s=a\nu$, and $\Pi_{L}(\nu)$ in eq. (\ref{pi_alpha<1}),
i.e.,
\begin{eqnarray}
P_\xi(G)
&=&
\int_0^\infty
p_{a\nu}(G)
\Pi_{L}(\nu)
{\rm d}\nu.
\end{eqnarray}
Now, using Eqs. (\ref{pi_alpha<1}) and (\ref{lngofl_a}) as well as the
scaling properties of the L\'evy distribution, we finally write the
distribution of conductances as
\begin{eqnarray}
\label{pofG_xi}
P_\xi(G)=\int_0^\infty p_{s(\alpha,\xi,z)}(G) q_{\alpha,1}(z){\rm d}z ,
\end{eqnarray}
where we have made again the change of variable
$
z=L/(2c\nu)^{1\over\alpha}
$
and defined
$
s(\alpha,\xi,z)={\xi}/(2{z^\alpha I_\alpha)}.
$
We remark that our result for $P_\xi(G) $ [eq. (\ref{pofG_xi})] depends
only on two parameters $\langle -\ln G \rangle_L =\xi$ and $\alpha$, as
we have previously announced. Thus, other details of the disorder
configurations are irrelevant.

\begin{figure}
 \includegraphics[width=\columnwidth]{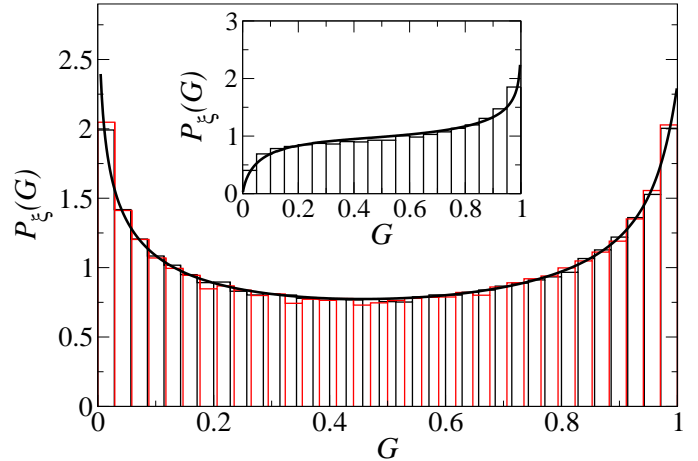}
\caption{(Color online) Conductance distributions from numerical
simulations (histograms) and theory (solid line), eq. (\ref{pofG_xi}).
Main frame: two histograms in black and red are shown (with different
size bins for a better distinction). Each histogram is generated from
different disorder distributions, $\rho_1(x)$ and $\rho_2(x)$
(see footnote 2), both with $\alpha=1/2$. In both cases $\xi=1.1$.
$P_\xi(G)$ shows two peaks at $G\sim 0$ and $G\sim 1$, revealing the
coexistence of insulating and ballistic regimes. Inset: $P_\xi(G)$ for
$\alpha=3/4$ with $\xi=0.76$ is shown. A good agreement between theory
and numerics is seen in all cases.}
\label{fig_3}
\end{figure}

Next we show some specific examples of the distribution of conductances
as given by eq.  (\ref{pofG_xi}). The theoretical results are compared
with numerical simulations using the model of potential barriers
introduced previously.  First we illustrate the fact that $P_\xi(G)$ is
only determined by the values of $\xi$ and $\alpha$. With this purpose
we have calculated numerically the conductance distributions from two
differently generated ensembles of wires: the value of $\xi$ is the {\it
same} in both cases, as well as the value of $\alpha$, but the
disorder-configurations of each ensemble follow {\it different density
distribution} \footnote[2]{Density distributions used for the numerical
simulations: $\rho_1(x)=q_{1/2,c}(x)=(c/\sqrt{2\pi})x^{-3/2}\exp{(-c^2/2x})$,
$\rho_2(x)=2^{-1}(1+x)^{-3/2}$
}. We thus expect the same conductance
distribution for both cases. This is shown in fig.  \ref{fig_3} where
the distribution $P_\xi(G)$ for each case (histograms in black and red)
are statistically the same. In addition, in  fig. \ref{fig_3},
we plot in solid line our theoretical result, eq. (\ref{pofG_xi}). A
good agreement with the numerical simulations is seen. On the other
hand, we point out that the conductance distributions exhibit two peaks,
at $G\sim 0$ and $G\sim 1$. This is an unconventional behavior in 1D
disordered systems and reveals the coexistence of insulating and
metallic (ballistic) regimes. Interestingly, a similar behavior of the
conductance distribution has been found in the so-called random-mass
Dirac model \cite{steiner}, where anomalously localized electronic
states \cite{garcia} might occur due to rare trapping disorder configurations. In the
inset of fig. \ref{fig_3} we show a further example of the conductance
distribution for wires with L\'evy-type disorder with $\alpha =3/4$.
Again, the agreement between theory and numerics is very good.

\begin{figure}
\includegraphics[width=\columnwidth]{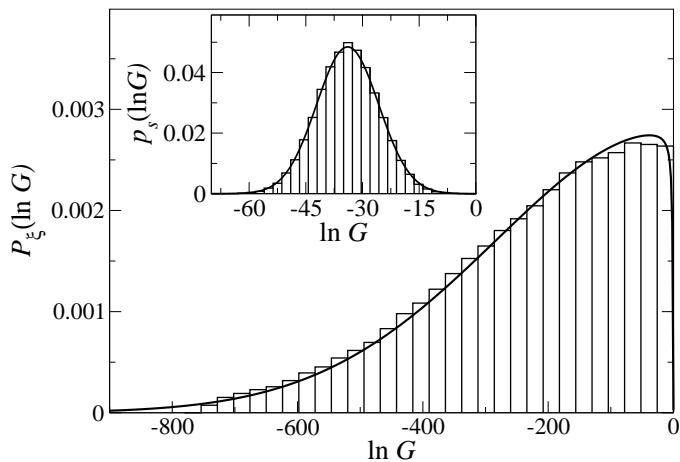}
\caption{Conductance distribution for large disordered wires with
$\alpha =1/2$ and $\xi=226$. We can see that $P_\xi(\ln G)$ do not
follow a Gaussian distribution, in contrast to the scaling theory
prediction for the insulating regime, like the example in the inset. A
good agreement is seen between analytical (solid line) and numerical
(histogram) results. }
\label{fig_4}
\end{figure}

Finally, we show an example of $P_\xi(G)$ in the insulating regime,
i.e., for very long wires. From the standard scaling theory, it is known
that $p_s(G)$ follows a log-normal distribution whose variance is twice
the mean value $\langle -\ln G \rangle$. An example of this standard
case is shown in the inset of fig. \ref{fig_4}. For long-range
disorder-configurations the situation is different. In the insulating regime, the
integral in eq. (\ref{pofG_xi}) can be performed analytically using
saddle point approximation. We plot the resulting distribution in fig.
\ref{fig_4} (main frame) for $\alpha=1/2$ and the corresponding result
from the numerical simulation (histogram). We can see clearly that
$P_\xi(\ln G)$ is not Gaussian, as we have remarked previously. Finally,
we point out that in all previous comparisons between theory and
simulations there are no free-fitting parameters: given the value of the
exponent $\alpha$, all we need is $\langle -\ln G \rangle_L (=\xi)$,
which we extract from the numerical experiments.

{\it Conclusions.--} L\'evy processes have been found in very different
phenomena in nature, but recent transport experiments indicate the
possibility of studying these processes in a controllable manner.
Motivated by these experimental achievements, we have studied the
statistical properties of the conductance across one-dimensional quantum
wires whose disorder configurations are characterized by probability
densities with a power-law tail, i.e., L\'evy-type disorder. We have
shown that the conductance distribution is completely determined by only
two parameters: the average $\langle \ln G \rangle_L$ and the scaling
exponent $\alpha$ of the power-law tail. Therefore the distribution of
conductances is universal in the sense that all wires with the same
$\langle \ln G \rangle_L$, but different disorder distributions obeying
the same scaling exponent $\alpha$ have the same conductance statistics.
Since we provide the complete conductance distribution other quantities
of interest such as the moments of the conductance and the shot-noise
power can be calculated from our results. We have shown that for $\alpha
\le 1$ the strong fluctuations of the conductance allow for the coexistence
of localized and delocalized regimes and the fluctuations of $\ln G$ do
not exhibit self-averaging. For $\alpha > 1$ the logarithm of the
conductance is a self-averaged quantity, as in the standard scaling
theory.
We think that the experimental observation of the different phenomena described
here such as the universality of the conductance statistics, in optical
materials or electronic transport experiments, should be of considerable
interest. Finally, we leave for future investigations the
extension of our results to higher-dimensional wires.

\acknowledgments
We acknowledge support from the MICINN (Spain) under Projects
FIS2009-07277 and FPA2009-09638.


\begin{thebibliography}{50}

\bibitem{anderson}
\Name{Anderson P. W., Thouless D. J., Abrahams E. \and Fisher D. S.}
\REVIEW{Phys. Rev. B}{22}{1980}{3519}.

\bibitem{fishes}
\Name{Sims D. W., et. al.}
\REVIEW{Nature}{45}{2008}{1098}.

\bibitem{stock}
\Name{Mantegna R. N. \and Stanley H. E.}
\REVIEW{Nature}{376}{1995}{46}.

\bibitem{levy-kolmorogov-calvo}
\Name{L\'evy P.}
\Book{Th\'eorie de l'addition des variables al\'eatoires},
\Publ{Gauthiers-Villars, Paris}
\Year{1937};
\Name{Gnedenko B. V. \and Kolmogorov A. N.}
Book{Limit distributions for sums of independent random variables}
\Publ{Addison-Wesley, Cambridge}
\Year{1954};
\Name{Calvo I., Cuch\'i J. C., Esteve J. G. \and Falceto F.}
\REVIEW{ J. Stat. Phys. }{141}{2010}{409}.

\bibitem{uchaikin}
\Name{Uchaikin V. V. \and V. M. Zolotarev V. M.}
\Book{Chance and Stability. Stable Distributions and their Applications}
\Publ{VSP, Utrecht, Netherlands, and references therein}
\Year{1999}

\bibitem{hideo}
\Name{Kohno H. \and Yoshida H.}
\REVIEW{Solid State Commun}{132}{2004}{59}

\bibitem{barthelemy}
\Name{Barthelemy P., Bertolotti J. \and Wiersma D. S.}
\REVIEW{Nature}{453}{2008}{495}.

\bibitem{lambert-boose-raffaella-beenakker-malyshev}
\Name{Leadbeater M., Falko V. I. \and Lambert C. J.}
\REVIEW{Phys. Rev. Lett.}{81}{1998}{1274};
\Name{Boos\'e D. \and Luck J. M.}
\REVIEW{J. Phys. A: Math. Theor.}{40}{2007}{140405};
\Name{Burioni R., Caniparoli L. \and
A. Vezzani} arXiv:1003.2161;
\Name{Beenakker C. W. J., Groth C. W., \and Akhmerov A. R.}
\REVIEW{Phys. Rev. B}{79}{2009}{024204};
\Name{A. Eisfeld, S. M. Vlaming S. M.,
Malyshev V. A. \and Knoester J.}
\REVIEW{Phys. Rev. Lett.}{105}{2010}{137402}

\bibitem{melnikov-dorokhov}
\Name{Mel'nikov V. I.}
\REVIEW{Pis'ma Zh. Eksp. Teor. Fiz.}{32}{1980}{244} [JETP Lett, {\bf 32}, 225, (1980)];
\Name{Dorokhov O. N.},
\REVIEW{Pis'ma Zh. Eksp. Teor. Fiz.}{36}{1982}{259} [JETP Lett,
{\bf 36}, 318, (1980)].

\bibitem{mello_groups}
\Name{Mello P. A.}
\REVIEW{J. Math. Phys.}{27}{1986}{2876}.

\bibitem{carlo-review}
\Name{Beenakker C. W. J.}
\REVIEW{Rev. Mod. Phys.}{69}{1997}{731}.

\bibitem{numericalgeneration} The incidence energy  of electrons is five times the barriers height. For numerical generation of L\'evy random
variables see Ref. \cite{uchaikin} and
\Name{Chambers J. M., Mallows C. L., \and
Stuck B. W.}
\REVIEW{J. Am. Stat. Ass}{71}{1976}{340}.

\bibitem{elsewhere} The results for  $1 < \alpha <2$ will be
published elsewhere.

\bibitem{steiner}
\Name{Steiner M., Chen Y., Fabrizio M. \and Gogolin A. O.}
\REVIEW{Phys. Rev. B}{59}{1999}{14848}.

\bibitem{garcia}
\Name{Garc\'ia-Garc\'ia A. M. \and Wang J.}
\REVIEW{Phys.  Rev.  Lett.} {100}{2008}{070603}


\end{thebibliography}
\end{document}